\documentclass[10pt,conference]{IEEEtran}
\usepackage{epsfig}
\usepackage{times}
\usepackage{float}
\usepackage{afterpage}
\usepackage{amsmath}
\usepackage{amstext}
\usepackage{amssymb,bm}
\usepackage{latexsym}
\usepackage{color}
\usepackage{array}
\usepackage{makecell}
\usepackage{diagbox}
\usepackage{amsmath}
\usepackage{amsthm}
\usepackage{graphicx}
\usepackage[center]{caption}
\usepackage{pstricks}
\usepackage{subfigure}
\usepackage{booktabs}
\usepackage{enumerate}
\usepackage{algorithm,algpseudocode}

\usepackage[left = 0.7in, right=0.7in, top=0.75 in, bottom=0.75in]{geometry}

\usepackage[normalem]{ulem}
\usepackage{url}
\newtheorem{theorem}{Theorem}

\newcommand{\Ksf}{\mathsf{K}} %number of users
\newcommand{\Qsf}{\mathsf{Q}} %number of layers
\newcommand{\esf}{\mathsf{e}} %erasure symbol
\newcommand{\Nbf}{\mathbf{N}} %channel state vector

\newcommand{\Scal}{\mathcal{S}}

\makeatletter
\def\BState{\State\hskip-\ALG@thistlm}
\makeatother

\begin{document}
%
% paper title
% can use linebreaks \\ within to get better formatting as desired
\title{%On The Capacity of 
%Inter User\&Layer Network Coded Retransmissions for
On the Capacity Region of the Layered Packet Erasure Broadcast Channel with Feedback}
%
%
% author names and IEEE memberships
% note positions of commas and nonbreaking spaces ( ~ ) LaTeX will not break
% a structure at a ~ so this keeps an author's name from being broken across
% two lines.
% use \thanks{} to gain access to the first footnote area
% a separate \thanks must be used for each paragraph as LaTeX2e's \thanks
% was not built to handle multiple paragraphs
%

\author{Siyao Li, Daniela Tuninetti, and Natasha Devroye\\
University of Illinois at Chicago, Chicago, IL 60607, USA\\
Email: \{sli210, danielat, devroye\}@uic.edu}

\maketitle

\begin{abstract}
In this paper the capacity region of the Layered Packet Erasure Broadcast Channel (LPE-BC) with Channel Output Feedback (COF) available at the transmitter is investigated. The LPE-BC is a high-SNR approximation of the fading Gaussian BC recently proposed by Tse and Yates, who characterized the capacity region for any number of users and any number of layers when there is no COF. %an open problem. 
This paper derives capacity inner and outer bounds for the LPE-BC with COF for the case of two users and any number of layers. The inner bounds generalize past results for the two-user erasure BC, which is a special case of the LPE-BC with COF with only one layer. The novelty lies in the use of \emph{inter-user \& inter-layer network coding} retransmissions (for those packets that have only been received by the unintended user), where each random linear combination may involve packets intended for any user originally sent on any of the layers.
%Various of inner bounds for two receivers are derived subject to their corresponding transmitting scheme. 
Analytical and numerical examples show that the proposed outer bound is optimal for some LPE-BCs.
% is given to evaluate the performance of different achievable schemes.
\end{abstract}

% IEEEtran.cls defaults to using nonbold math in the Abstract.
% This preserves the distinction between vectors and scalars. However,
% if the journal you are submitting to favors bold math in the abstract,
% then you can use LaTeX's standard command \boldmath at the very start
% of the abstract to achieve this. Many IEEE journals frown on math
% in the abstract anyway.

% Note that keywords are not normally used for peerreview papers.

% For peer review papers, you can put extra information on the cover
% page as needed:
% \ifCLASSOPTIONpeerreview
% \begin{center} \bfseries EDICS Category: 3-BBND \end{center}
% \fi
%
% For peerreview papers, this IEEEtran command inserts a page break and
% creates the second title. It will be ignored for other modes.
\IEEEpeerreviewmaketitle

\section{Introduction}
\label{sec:intro}
The Broadcast Channel (BC) is widely used as a model for downlink communication systems. 
A channel particularly important in wireless communications is the Additive White Gaussian Noise fading BC (AWGN-BC), where the channel between the single transmitter or base-station sending signal $X$, and multiple users is modeled as $Y_i = h_i X + N_i$ for user $i$, where $N_i$ is the AWGN, and $h_i$ is the fading parameter, or channel state.  When the transmitter has %an independent private message to transmit to each user, %and when the states $[h_1, h_2, \cdots h_k]$ are known at the transmitter and/or fixed, 
independent messages to send to different subsets of users, the capacity region, the largest set of rates for which the probability of error vanishes to zero as the blocklength increases to infinity, captures some of the tension seen in BCs: a single signal must be encoded such that when correlated versions of this signal are received at the users, each can extract their own intended message(s). %(possibly a common one as well) 
%from this single signal. 

%
%For wireless communications, the Additive White Gaussian Noise Broadcast Channel (AWGN-BC) is a relevant channel model. In this model,  at each channel use, the receiver $i$ receives $Y_i = h_i X + N_i$, for $X$ the transmitted signal, $N_i$ is the Gaussian noise at receiver $i$, and $h_i$ is the channel gain, or channel state between the transmitter and receiver $i$ (which may or may not change over time).  In general, the noise is assumed to be independent over time, forming a memoryless channel. 

While the capacity region of the general BC remains unknown, it is known for 
the degraded BC, %~\cite{cover_broadcast, cover_BC};
 %the capacity region of 
the BC with degraded message sets, %~\cite{Korner-Marton}; 
 %the capacity region of 
the AWGN-BC without fading, %~\cite{weingarten};
 %the capacity region of
and the AWGN-BC with fading known at the transmitter and the receivers~\cite{Cover:2006}. %~\cite{Li-Goldsmith}.
The capacity of the AWGN-BC with COF is unknown, but it may be enlarged by feedback even in the non-fading regime~\cite{Ozarow:1984,4655434Bhaskaran},  in sharp contrast to memoryless point-to-point channels. However, feedback cannot enlarge the capacity of the physically degraded BC~\cite{ElGamal-BC}. 

The capacity region of the AWGN-BC remains open when the fading / Channel State Information (CSI) is not available at the transmitter. % However, once the assumption that the transmitter has knowledge of the CSI,  the problem becomes open. 
Recently, the Layered Packet Erasure Broadcast Channel (LPE-BC) was proposed in~\cite{FadingBC} to approximate the AWGN-BC without transmitter CSI. %and determine its capacity region, and bound that of the Gaussian . 
%\nrd{FIX! Do not need all of it... in words}
In the LPE-BC, the base-station at each %scheduling slot (duration equal to the coherence slot of the fading channel)
channel use sends a vector of inputs (or layers of packets). At each time, each receiver receives a random number of layers, and missing layers are said to be ``erased''. Erasures are correlated because when a layer is erased, all the layers with smaller indices are also erased. The authors in \cite{FadingBC} determined the capacity region of the LPE-BC exactly and bounded that of the AWGN-BC to within a constant gap of $\approx6$ bits per channel use regardless of the fading distribution.

%
% the vector input $X^q := [X_1, X_2, \cdots, X_q]$ -- here $X_i$ is a ``packet'' that can be received if the SNR is within a certain range, and $q$ relates to the largest average SNRs in the fading AWGN counterpart~\cite{Tse-Yates}. The input $X^q$ is received at user $i$ as $Y_i := [X_1, X_2, \cdots X_{N_i}]$, where $N_i$ is a random variable on $\{0,1,\cdots q\}$ that characterizes the instantaneous channel strength and shows how many layers / packets %$X_i$ (each a fixed number of bits) 
%have been received by user $i$; notice that $Y_i$ does not contain packets $[X_{N_i+1}, \cdots, X_{q}]$ which are said to have been ``erased.'' Erasures are correlated because when a layer is erased, all the layers with lesser index are also erased. For this LPE-BC, the capacity region was derived exactly. 
%\nrd{FIX!!}For the LPE-BC, where the channel states corresponds to the $N_{k,t}$ number of layers received by user $k\in[\Ksf]$ at channel use $t\in[n]$, the following capacity region results are available or easily derived.

The LPE-BC also generalizes another channel model widely used in the networking literature: the (single-layer) Binary Erasure Channel (BEC-BC), where at each channel use a packet is sent, and the packet is either received or erased at each receiver. %(where the erasures across users may be correlated). 
The capacity region of the BEC-BC without COF is known for any number of users (i.e., because the channel is stochasticaly degraded)~\cite{Cover:2006}. 
%with COF for the case of two users~\cite{Sagduyu-unusual}. 
For the BEC-BC, the presence of COF allows the transmitter to know if a packet was erased or not at each receiver. 
This information allows it to re-send certain packets, and may do so in a {\it network-coded fashion} (by sending linar combinations of packets intended for different users). In~\cite{Sagduyu-unusual} the authors characterized the capacity region of a 2-user BEC-BC with COF and constructed several algorithms -- that employ network coding of packets received at the un-intended receiver -- that achieve this capacity. In~\cite{on-the-capacity-of-1-to-k}, the capacity region for 3-user BEC-BC as well as two types of symmetric $K$-user PEBCs and spatially independent PEBCs with one-sided fairness constraints with COF were derived. Similar results to~\cite{on-the-capacity-of-1-to-k} were also obtained in~\cite{multiuser-BEC-with-feedback}.
%An outer bound to the capacity region of $K$-user PEBCs with COF was computed in~\cite{multiuser-BEC-with-feedback} where algorithms were presented that achieve this outer bound for 3 users under the condition that {\blue$\mathcal{R}_{\text{ord}} \supseteq \mathcal{C}^{\text{out}}$ in Theorem 2, or arbitrary channel statistics for $N=3$.} %\nrd{for the case of  3 users.}

%The LPE-BC consists of $q$ layers ordered from the most significant to less significant layers, in which erasures are correlated,  and the channel is not limited to degraded. Fading is modeled as erasures of the less significant layers.

%\subsection{Preliminary work: capacity regions of the $\Ksf$ user LPE-BC}
\paragraph*{Contributions}
All exact capacity results for the LPE-BC are without COF~\cite{FadingBC}, or for the single-layer case with COF and up to three users~\cite{Sagduyu-unusual,on-the-capacity-of-1-to-k,multiuser-BEC-with-feedback}. We look explicitly at the (multi-layer) LPE-BC with COF and combine and extend the work in~\cite{FadingBC} and~\cite{Sagduyu-unusual,on-the-capacity-of-1-to-k,multiuser-BEC-with-feedback}. We provide a general outer bound for LPE-BC with COF for $\Ksf$ receivers ($\Ksf\geq 2$) and $\Qsf$ layers ($\Qsf\geq 1$), and present several achievable rate regions (some only for the 2-user case). These regions are obtained using schemes that employ network coding per-layer and / or across layers in case retransmissions are needed. Inner and outer bounds are analytically and numerically compared; it is seen that they meet for certain LPE-BCs, thus giving exact capacity results.

%The presented inner bounds and differs from the previous works in that, although it also employs the thought of transmitting linear-combination packets, it interpolates additional concepts and algorithms that generalize the results considering achievable rates to multiple layers. In particular, we present the capacity region of the LPE-BC  and full foresee transmitter CSI. The outer bound of LPE-BC with COF for $K$ receivers ($K\geq 2$) is proved as well. Moreover, several algorithms and the corresponding inner bounds are proposed and analyzed. 

\paragraph*{Paper Organization}
Section~\ref{sec:model} introduces the LPE-BC; 
Section~\ref{sec:schemes} presents the information theoretic inner and outer bounds to the capacity region of the  LPE-BC with COF known at transmitter; 
Section~\ref{sec:numerical} illustrates the derived achievable regions and outer bound with a numerical example; Section~\ref{sec:concl} concludes the paper.

\section{System Model and Ergodic Capacity Results} 
\label{sec:model}
The LPE-BC, as originally proposed in~\cite{FadingBC}, consists of one transmitter (base-station) and $\Ksf$ receivers (users).
At each channel use ({\it slot}) the transmitter sends $\Qsf$ symbols ({\it packets / layers}), each symbol from an input alphabet $\mathcal{X}$, where $\mathcal{X}$ is assumed to be a discrete finite set; 
the input is denoted as $X^{\Qsf} := (X_{1}, \ldots, X_{\Qsf}) \in \mathcal{X}^{\Qsf}$.
The LPE-BC is characterized by the random vector ({\it channel state}) $\Nbf := (N_{1}, \ldots, N_{\Ksf})\in[0:\Qsf]^{\Ksf}$, 
where $N_{k}$ denotes how many layers have been successfully received by user $k\in[\Ksf]$.
The LPE-BC channel output for user $k\in[\Ksf]$ is $Y_{k} := X^{N_{k}} = (X_{1}, \ldots, X_{N_{k}})$ for $N_{k}>0$, that is, layers $(X_{N_{k}+1}, \ldots, X_{\Qsf})$ have been {\it erased}; if $N_{k}=0$ then all layers have been erased and we set $Y_{k}=\esf$ for some constant $\esf$. The channel state $\Nbf$ is assumed to be independent and identically distributed (i.i.d.) across time slots, that is, the channel is memoryless. In the LPE-BC, the erasures are correlated so as to capture the high SNR behavior of the fading AWGN-BC~\cite{FadingBC}.
The case $\Qsf=1$ and $\mathcal{X}=$GF(2) is the well studied BEC-BC.

A code for the LPE-BC is defined as follows.
The transmitter must convey $|\mathcal{X}|^{n R_{k}}$ (private) messages reliably to user $k\in[\Ksf]$ in $n$ channel uses. Note that the rate $R_{k}$ is measured in number of packets per channel use. Let $(W_{1},\ldots,W_{\Ksf})$ be the messages to be sent to the users. We distinguish different cases %(encoding functions) 
based on the amount of CSI at the transmitter (CSIT):
\begin{enumerate}

\item \label{item:nof} no CSIT:
$X^{\Qsf}_{t}(W_{1},\ldots,W_{\Ksf}), \ t\in[n],$

\item \label{item:cof} COF:
$X^{\Qsf}_{t}(W_{1},\ldots,W_{\Ksf}, \Nbf^{t-1}), \ t\in[n],$

%\item \label{item:currrent} current CSIT:
%$X^{\Qsf}_{t}(W_{1},\ldots,W_{\Ksf}, \Nbf^{t}), \ t\in[n],$

\item \label{item:fla} full-lookahead CSIT:
$X^{\Qsf}_{t}(W_{1},\ldots,W_{\Ksf}, \Nbf^{n}), \ t\in[n],$

\end{enumerate}
where $X^{\Qsf}_{t}(\cdot)$ is the encoding function a time $t$.
We assume that all receivers have full CSI, namely, by time $t=n$ they know $\Nbf^{n}$.
User $k\in[\Ksf]$ estimates $\widehat{W}_{k} = {\rm dec}_{k}(Y_{k}^{n},\Nbf^{n})$ for some decoding function ${\rm dec}_{k}$. The probability of error is $P_e^{(n)} := 1-\Pr[{\rm dec}_{k}(Y_{k}^{n},\Nbf^{n})=W_{k}, \ \forall k\in[\Ksf]]$.
The capacity region is the convex closure of the set of $(R_{1}, \ldots, R_{\Ksf})\in\mathbb{R}^{\Ksf}_{+}$ that can be decoded at the receivers with vanishing probability of error for some blocklength $n$, i.e., $\lim_{n\to\infty} P_e^{(n)} = 0.$

The case in item~\ref{item:nof} / no CSIT has been solved in~\cite{FadingBC}:
\begin{theorem}[no CSIT: from~\cite{FadingBC}]
\label{thm:Tse-Yates ITA-2011}
The capacity region of the LPE-BC with no CSIT is characterized by %completely 
\begin{align}
\sum_{k\in[\Ksf]} \omega_{k} R_{k} \leq \sum_{q\in[\Qsf]} \max_{u\in[\Ksf]}\left( \omega_{u} \Pr[N_{u} \geq q] \right), 
\label{eq:thm:Tse-Yates ITA-2011}
\end{align}
for all $(\omega_{1}, \ldots, \omega_{\Ksf})\in\mathbb{R}^{\Ksf}_{+}$.%\omegabf := 
\end{theorem}
In this paper we are interested in the capacity for the case in item~\ref{item:cof}.
%The case in item~\ref{item:currrent} seems to be as the one in item~\ref{item:fla}.
%
The case in item~\ref{item:fla} is trivially solved by:
\begin{theorem}[full-lookahead CSIT / ergodic capacity]
\label{thm:fla-trivial}
The capacity region of the LPE-BC with full lookahead CSIT is characterized by %completely 
%\vspace{-0.1cm}
\begin{align}
\sum_{k\in\Scal} R_{k} \leq  \mathbb{E}[\max(N_u : u\in\Scal)],
\label{eq:thm:fla-trivial} 
\end{align}
%\vspace{-0.1cm}
for all non-empty subsets $\Scal \subseteq [\Ksf]$.
\end{theorem}
%The rate point 
%\begin{align*}
%R_{k}
%& = \mathbb{E}[\max(N_{1},\ldots,N_{k})]-\mathbb{E}[\max(N_{1},\ldots,N_{k-1})]
%\\
%& = \mathbb{E}\left[ \left[ N_{k}-\max(N_{1},\ldots,N_{k-1})\right]^+ \right],
% \forall k\in[\Ksf]
%\end{align*}
%can be achieved as follows:
%in each slot $t$, the transmitter knows $\Nbf_{t}$;
%it assigns the top $N_{1,t}$ layers to user~1,
%then it assigns the subsequent $[N_{2,t}-N_{1,t}]^+$ layers to user~2,
%then it assigns the subsequent $[N_{3,t}-\max(N_{1,t},N_{2,t})]^+$ layers to user~3,
%etc; this achieves the claimed rate point. 
%All other corner points can be achieved similarly, just consider an appropriate permutation on the user indices.

\section{Capacity of the LPE-BC with COF} %: $X^{\Qsf}_{t}(W_{1},\ldots,W_{\Ksf},\Nbf^{t-1}), t\in[n]$}
\label{sec:schemes}
Although COF does not increase the capacity of a memoryless single user channel, it enlarges the capacity region of broadcast channels in general~\cite{Ozarow:1984,4655434Bhaskaran}. 

%Herein, we propose a general outer bound of the LPE-BC with COF for $\Ksf$ users and $\Qsf$ layers. Several inner bounds are derived in terms of the corresponding transmitting schemes when $\Ksf = 2$. %multiuser channels in general~\cite{ADD}. 

\paragraph*{Outer Bound}
%\par{Outer Bound}:
The following theorem gives an outer bound to the capacity of the LPE-BC with COF:

\begin{theorem}[COF: new outer bound]
\label{thm:COFnewOuter}
The capacity region of the LPE-BC with COF is contained into
\begin{align}
&\sum_{k\in[\Ksf]} \omega_{k} R_{k} 
  \leq 
  \sum_{q\in[\Qsf]} \max_{k\in[\Ksf]}\left( \omega_{\pi(k)} \Pr[\max(
 N_{\pi(k)}^{\pi(\Ksf)}
 %N_{\pi(k)},N_{\pi(k+1)},\ldots,N_{\pi(\Ksf)}
 ) \geq q] \right),
\label{eq:converse thm:COFnewOuter}
\end{align}
for all $(\omega_{1}, \ldots, \omega_{\Ksf})\in\mathbb{R}^{\Ksf}_{+}$  and for all permutations $\pi$ of $[\Ksf]$, and where $N_{\pi(k)}^{\pi(\Ksf)} := [N_{\pi(k)},N_{\pi(k+1)},\ldots,N_{\pi(\Ksf)}]$. 
%for all $\omegabf := (\omega_{1}, \ldots, \omega_{\Ksf})\in\mathbb{R}^{\Ksf}_{+}$.
\end{theorem}
\begin{IEEEproof}
%Converse: 
We enhance the original LPE-BC to a physically degraded LPE-BC %. We start by finding an upper bound to the capacity region with COF 
by using a cooperation-based argument; then, since feedback cannot increase the capacity of the physically degraded broadcast channel~\cite{ElGamal-BC}, for the found physically degraded LPE-BC we use the capacity result in Theorem~\ref{thm:Tse-Yates ITA-2011}.
\begin{subequations}
Consider a permutation $\pi$ of $[\Ksf]$. 
Enhance / give as genie side information to receiver $\pi(k)$ the following
\begin{align}
\widetilde{N}_{\pi(k)} := \max(N_{\pi(k)},N_{\pi(k+1)},\ldots,N_{\pi(\Ksf)}),
\end{align}
so that the following Markov chains hold
\begin{align}
X^{\Qsf} \to X^{\widetilde{N}_{\pi(1)}} \to X^{\widetilde{N}_{\pi(2)}} \ldots \to X^{\widetilde{N}_{\pi(\Ksf)}},
\\
X^{\Qsf} \to X^{\widetilde{N}_{k}} \to X^{N_{k}}, \ \forall k\in[\Ksf].
\end{align}
\label{eq:enhancement thm:COFnewOuter}
\end{subequations}
Apply Theorem~\ref{thm:Tse-Yates ITA-2011} to the enhanced LPE-BC in~\eqref{eq:enhancement thm:COFnewOuter} to obtain the region in~\eqref{eq:converse thm:COFnewOuter}.
\end{IEEEproof}

Note that Theorem~\ref{thm:COFnewOuter} with $\Qsf=1$ is the outer bound in~\cite{on-the-capacity-of-1-to-k} whose tightness is discussed next.

\paragraph*{Inner Bounds}
%Inner Bounds:
We give next several inner bounds for the LPE-BC with COF.

%\subsubsection{A trivial achievable region}
\begin{theorem}[COF: new Ach1]
\label{thm:scheme1}
The following region is achievable for the LPE-BC with COF and $\Ksf=2$ users:
\begin{subequations}
\begin{align}
&\{ (R_1,R_2) : \max_{q\in[\Qsf]} \left( v_{q} \right) \leq 1 \text{ for some } R_{u,q}\geq0 \},
\\
&v_{q} := \max\left(
% \frac{R_{1,q}}{1-\epsilon_{12,q}} + \frac{R_{2,q}}{1-\epsilon_{2,q}},
% \frac{R_{1,q}}{1-\epsilon_{1,q}}  + \frac{R_{2,q}}{1-\epsilon_{12,q}}
 \frac{R_{1,q}}{\Pr[\max(N_1,N_2) \geq q]} + \frac{R_{2,q}}{\Pr[N_2 \geq q]},
\right.\notag\\&\quad\left. 
 \frac{R_{1,q}}{\Pr[N_1 \geq q]}  + \frac{R_{2,q}}{\Pr[\max(N_1,N_2) \geq q]}
\right), \ q\in[\Qsf], 
\label{trivial_achievable_region}
%\\
%&
%1-\epsilon_{u,q} = \Pr[N_u \geq q], \ u\in[2], q\in[\Qsf],
%\\
%&
%1-\epsilon_{12,q}= \Pr[\max(N_1,N_2) \geq q],  \ q\in[\Qsf],
\\
&R_u := R_{u,1}+\ldots+R_{u,\Qsf}, \ u\in[2].
\end{align}
\label{eq:scheme1}
\end{subequations}
\end{theorem}
%\vspace{-0.5cm}
\begin{IEEEproof}
The region in~\eqref{eq:scheme1} is achievable for the LPE-BC by using the scheme in~\cite{Sagduyu-unusual} independently on each layer, where the erasure channel model studied in~\cite{Sagduyu-unusual} is the special case of $\Qsf=1$ in out LPE-BC model. To map the notation used in~\cite{Sagduyu-unusual} to ours, please note that $\epsilon_{u,q} = 1-\Pr[N_u \geq q], \ u\in[2], q\in[\Qsf]$  is the probability that layer $q$ is erased for user $u$, and
$\epsilon_{12,q}= 1-\Pr[\max(N_1,N_2) \geq q],  \ q\in[\Qsf]$ is the probability that layer $q$ is erased at both users.
\end{IEEEproof}

Note that the extension of Theorem~\ref{thm:scheme1} to more than $\Ksf =2$ users requires knowing the capacity of the single-layer model for $\Ksf$ users, which is open at present in general. The scheme in~\cite{on-the-capacity-of-1-to-k} is tight (i.e., it achieves the outer bound in Theorem~\ref{thm:COFnewOuter}) for $\Qsf=1$ and $\Ksf\leq 3$ users, and also for $\Qsf=1$ and $\Ksf\geq 4$ in some symmetric settings; the same paper claims that the scheme matches to numerical precision the outer bound for all simulated case of $\Ksf\leq 6$ users; if the scheme were indeed optimal for any number of users, then Theorem~\ref{thm:scheme1} could give a scheme for any number of layers and users, and would prove the tightness of Theorem~\ref{thm:COFnewOuter} for $\Qsf=1$.

%\subsubsection{Less trivial achievable regions}
For the rest of this section, the achievable regions for the LPE-BC with COF and $\Ksf = 2$ users will be of the form presented in Theorem~\ref{thm:Prototype} next, which was inspired by~\cite{Sagduyu-unusual}.
We shall use the following nomenclature:
an {\it uncoded packet} is packet that is sent by itself, i.e., not coded together with other packets, on some layer;
an {\it overheard packet} is packet that has not yet been delivered uncoded to the intended user but it has  been successfully received at the non-intended user; and
a {\it (network) coded packet} is packet that is sent on some layer in a linear combination involving other packets that were originally sent uncoded on possibly some other layer and to some other user.
The idea is to have a protocol with two phases:
Phase1 corresponds to uncoded transmission on some layers (and can be split in sub-phases), while
Phase2 to network coded transmissions on all layers.
\begin{theorem}[COF: new Ach2]
\label{thm:Prototype}
The following region is achievable for the LPE-BC with COF and $\Ksf=2$ users:
\begin{subequations}
\begin{align}
\mathcal{R}_\text{COF} &:= \{ (R_1,R_2) : t^\text{\rm(unc)} + t^\text{(NC)} \leq t 
\notag\\&\quad \text{ for some } t\geq0, k_{u,q}\geq 0, q\in[\Qsf], u\in[2] \},
\\
 R_u &:= \frac{\sum_{q\in[\Qsf]}k_{u,q}}{t}, 
\forall u\in[2], \ \text{(rate)},
\\
 t^\text{\rm(unc)} &:= \max_{q\in[\Qsf]} \left( t^\text{\rm(unc)}_{q} \right), \ \text{(duration of Phase1)},
\label{eq:schemePrototype-T1}
\\
 t^\text{\rm(unc)}_{q} &:= \frac{k_{1,q} + k_{2,q}}{\Pr[\max(N_{1},N_{2})\geq q]}, 
\forall q\in[\Qsf], 
\label{eq:schemePrototype-T1q}
\\
 k^\text{\rm(rem)}_{u,q} &:= k_{u,q}\left(1\!-\!\frac{\Pr[N_{u}\geq q]}{\Pr[\max(N_{1},N_{2})\geq q]}\right),
\!\!\!\!\!
\begin{array}{l}
\forall q\in[\Qsf], \\
\forall u\in[2], \\
\end{array}
\label{eq:schemePrototype-Kuqremaining}
\\
 t^\text{(NC)} &:= \max_{u\in[2]}\left( t^\text{(NC)}_{u} \right), \ \text{(duration of Phase2)},
\label{eq:schemePrototype-T2}
\\
 t^\text{(NC)}_{u} &:=  \frac{k^\text{\rm(rem)}_{u}}{\mathbb{E}[N_{u}]},
\forall u\in[2], 
\label{eq:schemePrototype-T2u}
\\
 k^\text{\rm(rem)}_{u} &:= \text{\rm``DEPENDS ON THE SCHEME''}, \ \forall u\in[2].
\label{eq:schemePrototype-KuNotDefined}
\end{align}
\label{eq:schemePrototype}
\end{subequations}
\end{theorem}
%\vspace*{-.8cm}
\begin{IEEEproof} %\in[\Qsf]
%The idea for the region in~\eqref{eq:schemePrototype} is to have a two-phase scheme. 
Let $k_{u,q} \gg 1, \ u\in[2], \ q\in[\Qsf]$ so that we can invoke the Law of Large Numbers in the following analysis (loosely speaking, we ``replace'' random processes with their statistical averages).

In Phase1, we send $k_{u,q}$ uncoded packets on layer $q\in[\Qsf]$ for user $u\in[2]$, one by one until one of the two users has received it; it takes on average $\frac{1}{\Pr[\max(N_{1},N_{2})\geq q]}$ time slots to deliver one uncoded packet to some user on layer $q\in[\Qsf]$, and therefore layer $q\in[\Qsf]$ is done delivering all its uncoded packets by time $t^\text{\rm(unc)}_{q}$ in~\eqref{eq:schemePrototype-T1q} at which point the number of overheard packets for user $u\in[2]$ is $k^\text{\rm(rem)}_{u,q}$ in~\eqref{eq:schemePrototype-Kuqremaining}.
By time $t^\text{\rm(unc)}$ in~\eqref{eq:schemePrototype-T1} all layers are done sending uncoded packets and there are $k^\text{\rm(rem)}_{u}$ in~\eqref{eq:schemePrototype-KuNotDefined} packets that still need to be delivered to user $u\in[2]$, which can be sent coded on any layer.

In Phase2, once all layers are done sending their uncoded packets at time $t^\text{\rm(unc)}$ in~\eqref{eq:schemePrototype-T1}, we send  on every layer different linearly independent random linear combinations of the overheard packets;
% (there are in total $\sum_{u\in[2]}k^\text{\rm(rem)}_{u}$ such packets); this phase is network coded;
user $u\in[2]$ receives on average $\mathbb{E}[N_{u}]$ packets in each time slot, thus it is done receiving its remaining $k^\text{\rm(rem)}_{u}$ in~\eqref{eq:schemePrototype-KuNotDefined} packets in $ t^\text{(NC)}_{u}$ in~\eqref{eq:schemePrototype-T2u} time slots.

The different schemes in the following differ in the way the time slots in the interval $[t^\text{\rm(unc)}_{q},t^\text{\rm(unc)}]$ on layer $q\in[\Qsf]$ are utilized; this is the time interval after which all the $k_{1,q} + k_{2,q}$ uncoded packets for layer $q\in[\Qsf]$ have been delivered to at least one user but there is at least one layer that is not yet done sending its uncoded packets. Possible choices are to leave layer $q\in[\Qsf]$ idle during $[t^\text{\rm(unc)}_{q},t^\text{\rm(unc)}]$ or to start sending some coded packets.
%Note that the overheard packets-not-yet-delivered-to-user$u$-but-available-at-user$\bar{u}$ on layer $q$ are given by $k^\text{\rm(rem)}_{u,q}$ in~\eqref{eq:schemePrototype-Kuqremaining} with $u\in[2], \bar{u}\in[2], \bar{u}\not= u$, which can be sent by using random network coding.
\end{IEEEproof}

Next we propose various ways to transmit information on a layer once its uncoded phase if over, this will give different expressions for the term in~\eqref{eq:schemePrototype-KuNotDefined} in Theorem~\ref{thm:Prototype}.

\begin{theorem}[COF: new Ach2: a layer is idle once its uncoded phase is over]
\label{thm:scheme2}
The region in~\eqref{eq:schemePrototype} is achievable %for the LPE-BC with COF and $\Ksf=2$ users 
with $k^\text{\rm(rem)}_{u}$ in~\eqref{eq:schemePrototype-KuNotDefined} given by
\begin{align}
k^\text{\rm(rem)}_{u} 
&=\sum_{q\in[\Qsf]} k^\text{\rm(rem)}_{u,q}, \ \forall u\in[2],
\label{eq:scheme2-Ku}
\end{align}
for $k^\text{\rm(rem)}_{u,q}$ in~\eqref{eq:schemePrototype-Kuqremaining}.
\end{theorem}
\begin{IEEEproof}
Here nothing is sent on layer $q\in[\Qsf]$ during times slots $[t^\text{\rm(unc)}_{q},t^\text{\rm(unc)}]$, thus in Phase2 all the overheard packets from all layers have to be delivered as indicated by~\eqref{eq:scheme2-Ku}.
\end{IEEEproof}

Note that the extension of Theorem~\ref{thm:scheme2} to more than 2 users requires being able to track which subset of non-intended users has received a certain uncoded packet; this is the same stumbling block as in the single-layer case in~\cite{on-the-capacity-of-1-to-k} for $\Ksf\geq 4$.

\begin{theorem}[COF: new Ach2: a layer, once its uncoded phase is over, uses network coding for its overheard packets only]
\label{thm:scheme3-SL}
The region in~\eqref{eq:schemePrototype} is achievable %for the LPE-BC with COF and $\Ksf=2$ users 
with $k^\text{\rm(rem)}_{u}$ in~\eqref{eq:schemePrototype-KuNotDefined} given by
\begin{align}
&k^\text{\rm(rem)}_{u} 
\notag\\ &=
\sum_{q\in[\Qsf]} \Bigg[k^\text{\rm(rem)}_{u,q} 
-  (t^\text{\rm(unc)} - t^\text{\rm(unc)}_{q})\Pr[N_{u}\geq q] \Bigg]^+,  
 \forall \ u \in[2].
\label{eq:scheme3SL-Ku}
\end{align}
for $k^\text{\rm(rem)}_{u,q}$ in~\eqref{eq:schemePrototype-Kuqremaining},
$t^\text{\rm(unc)}$ in~\eqref{eq:schemePrototype-T1}
and $t^\text{\rm(unc)}_{q}$ in~\eqref{eq:schemePrototype-T1q}.
\end{theorem}
\begin{IEEEproof}
The region in Theorem~\ref{thm:scheme3-SL} is the following enhancement of Theorem~\ref{thm:scheme2}.
During Phase1 of Theorem~\ref{thm:scheme2}, layer $q\in[\Qsf]$ remains idle during $[t^\text{\rm(unc)}_{q},t^\text{\rm(unc)}]$, which is a clear waste of resources. The idea in Theorem~\ref{thm:scheme3-SL} is that as soon as a layer finishes sending its uncoded packets, it immediately starts sending network-coded overheard packets that need retransmission on that layer. The number of overheard packets for user $u\in[2]$ on layer $q\in[\Qsf]$ at time slot $t^\text{\rm(unc)}_{q}$ is $k^\text{\rm(rem)}_{u,q}$. There are extra $t^\text{\rm(unc)} - t^\text{\rm(unc)}_{q}$ time slots to transmit coded packets on layer $q\in[\Qsf]$ before the start of Phase2 (when all layers will send coded packets). The number of packets that can be received on layer $q\in[\Qsf]$ by user $u\in[2]$ is $k^\text{(extra)}_{u,q} = (t^\text{\rm(unc)} - t^\text{\rm(unc)}_{q})\Pr[N_{u}\geq q]$. Since user $u\in[2]$ has $k^\text{\rm(rem)}_{u,q}$ packets that still need to be received on layer $q\in[\Qsf]$, $k^\text{(extra)}_{u,q}$ can not be larger than $k^\text{\rm(rem)}_{u,q}$. 
%For the slowest layer $q$, it will not contribute extra XORed packets can be received by user $u$, and because $\min\left((t^\text{\rm(unc)} - t^\text{\rm(unc)}_{q})\Pr[N_{u}\geq q], k^\text{\rm(rem)}_{u,q} \right) = 0$, which makes the slowest layer is also taken into account. 
Thus, we have $k^\text{\rm(rem)}_{u}$ in~\eqref{eq:scheme3SL-Ku} packets left for user~$u$ when  Phase1 ends.
\end{IEEEproof}

The scheme in Theorem~\ref{thm:scheme3-SL} tries to ``fill'' the idle slots in the scheme in Theorem~\ref{thm:scheme2}. However, it may still be the case that once a layer is done sending linear combinations of its overheard packets, other layers are still in the process of completing their uncoded phases; when this is the case, this layer will remain idle, which does not seem to be optimal. The following scheme aims to eliminate all idle slots.

\begin{theorem}[COF: new Ach2: a layer, once its uncoded phase is over, sends coded packets by combining all overheard packets from all layers up to that point]
\label{thm:scheme3-DT}
The region in~\eqref{eq:schemePrototype} is achievable %for the LPE-BC with COF and $\Ksf=2$ users 
with $k^\text{\rm(rem)}_{u}$ in~\eqref{eq:schemePrototype-KuNotDefined} given by
\begin{align}
&k^\text{\rm(rem)}_{u}
\notag\\ &=  
\Bigg[ \sum_{q \in [\Qsf]} k_{u,q}^{\text{\rm(rem)}} - (t^\text{\rm(unc)} - t^\text{\rm(unc)}_{q})\Pr[N_{u}\geq q]\Bigg] ^+,  
\forall \ u \in[2]. 
\label{eq:scheme3DT-Ku}
\end{align}
%for $k^\text{(rtx)}_{u}[\Qsf]$ in~\eqref{eq:DTOct04proposed_k_rtx}.
\end{theorem}
\begin{IEEEproof}
The region in Theorem~\ref{thm:scheme3-DT} is the following enhancement of Theorem~\ref{thm:scheme3-SL}. 
During Phase1 of Theorem~\ref{thm:scheme3-SL}, once layer $q\in[\Qsf]$ has finished sending its uncoded packets at time $t^\text{\rm(unc)}_{q}$, we send linear combinations of the overheard packets on layer $q$ and the network coded packets are sent on layer $q$ only; we refer to this scheme as {\it inter-layer network coding scheme}.
In Theorem~\ref{thm:scheme3-DT} we propose an {\it inter-layer network coding scheme}: once layer $q$ has finished sending its uncoded packets at time $t^\text{\rm(unc)}_{q}$, we send linear combinations of *all* overheard packets on *all* layers up to time $t^\text{\rm(unc)}_{q}$ (note: each layer gets a linearly independent linear combination). %The layers that finish before layer $q$ are indexed by $\mathcal{F}_{q} := \{j\in[\Qsf] : T_{1,j} \leq t^\text{\rm(unc)}_{q}\}$. 

\begin{subequations}
Moreover, for Theorem~\ref{thm:scheme3-DT} the order in which packets are sent on layer $q\in[\Qsf]$ during the uncoded phase (that is, time interval $[0,t^\text{\rm(unc)}_{q}]$) is randomized, that is, the probability of a user being picked to be served at a given time slot is proportional to how many uncoded packets that user needs to receive on that layer. 
Let $A_q$ be the random variable that indicates which user is served on layer $q\in[\Qsf]$ during the uncoded phase, assumed to be i.i.d. over time and independent of everything else with
\begin{align}
&\Pr[A_q=u] = \frac{k_{u,q}}{k_{1,q}+k_{2,q}}, u\in[2].
\label{eq:pr Aq}
\end{align}
With~\eqref{eq:pr Aq}, we write
\begin{align}
&\Pr[A_q=u, \max(N_1,N_2) \geq q]
 = \frac{k_{u,q}}{t^\text{\rm(unc)}_{q}},
%= \frac{k_{u,q}}{\text{eq\eqref{eq:schemePrototype-T1}}};
\label{eq:pr Aq max}
\\
&\Pr[A_q=u, \max(N_1,N_2) \geq q, N_u<q] 
%= \frac{k_{u,q}}{t^\text{\rm(unc)}_{q}}\left(1-\frac{\Pr[N_u\geq q]}{\Pr[\max(N_1,N_2) \geq q]}\right)
 = \frac{k_{u,q}}{t^\text{\rm(unc)}_{q}} \ \eta_{u,q},
% = \frac{k^\text{\rm(rem)}_{u,q}}{t^\text{\rm(unc)}_{q}}
\label{eq:pr Aq max Nu}
\\
&%\text{ with } 
\eta_{u,q} := \frac{k^\text{\rm(rem)}_{u,q}}{k_{u,q}} = 1-\frac{\Pr[N_{u}\geq q]}{\Pr[\max(N_{1},N_{2})\geq q]}\in[0,1],
\label{eq:DTOct04proposed_eta_uq}
\end{align}
where~\eqref{eq:pr Aq max} is the probability that user $u\in[2]$ is scheduled on layer $q\in[\Qsf]$ and its uncoded packet is received by at least one of the users; 
similarly,~\eqref{eq:pr Aq max Nu} is the probability that user $u\in[2]$ is scheduled on layer $q\in[\Qsf]$ and its uncoded packet is received by the other user only. The quantity in~\eqref{eq:DTOct04proposed_eta_uq} can be thought of as the fraction of overheard packets for user $u\in[2]$ on layer $q\in[\Qsf]$.

Let $\pi$ be the permutation of $[\Qsf]$ such that 
\begin{align}
0\equiv t^\text{\rm(unc)}_{\pi(0)}\leq t^\text{\rm(unc)}_{\pi(1)} \leq t^\text{\rm(unc)}_{\pi(2)} \ldots \leq t^\text{\rm(unc)}_{\pi(\Qsf)} \equiv t^\text{\rm(unc)}.
\end{align} 
Let also 
\begin{align}
\Delta_j := t^\text{\rm(unc)}_{\pi(j)}-t^\text{\rm(unc)}_{\pi(j-1)}, \ j\in[\Qsf]. 
\end{align}
Phase1 is composed of $\Qsf$ sub-phases, where the $j$-th sub-phase has duration $\Delta_j$, i.e., time slots $[t^\text{\rm(unc)}_{\pi(j-1)},t^\text{\rm(unc)}_{\pi(j)}), \ j\in[\Qsf]$. At time $t^\text{\rm(unc)}_{\pi(j)}$, the layers $\pi(1), \ldots, \pi(j)$ have finished their uncoded phase. There are $\Qsf!$ possible configurations of sub-phases, one for each  permutation of $[\Qsf]$.

Let $k^\text{\rm(unc)}_{u,q}[j]$ be the number of uncoded packets left to be delivered to user $u\in[2]$ on layer $q\in[\Qsf]$ at the end of the $j$-th sub-phase; these packets must be still sent on layer $q\in[\Qsf]$.
Also, let $k^\text{(rtx)}_{u}[j]$ be the number of overheard packets left to be delivered to user $u\in[2]$ at the end of the $j$-th sub-phase; these packets can be sent coded on any layer.
Initialize $k^\text{\rm(unc)}_{u,q}[0] = k_{u,q} \geq 0$ and $k^\text{(rtx)}_{u}[0] = 0$.
We have the following recursive equation for $j\in[\Qsf]$:
\begin{align}
   &k^\text{\rm(unc)}_{u,q}[j]  \nonumber \\
   &= \left[ k^\text{\rm(unc)}_{u,q}[j-1] - \Delta_j \Pr[A_q=u, \max(N_1,N_2) \geq q] \right]^+
\notag\\ &= k_{u,q} \max\left( 1-\frac{t^\text{\rm(unc)}_{\pi(j)}}{t^\text{\rm(unc)}_{q}}, \ 0 \right).
\label{eq:DTOct04proposed_k_unc}
\end{align}
The update equation for $k^\text{\rm(unc)}_{u,q}[j]$ in~\eqref{eq:DTOct04proposed_k_unc} says that the number of uncoded packets for user $u\in[2]$ on layer $q\in[\Qsf]$ decreases with ``time'' $j\in[\Qsf]$. In particular, at the end of the $j$-th sub-phase, $k^\text{\rm(unc)}_{u,q}[j-1]$ is reduced by the number of packets that can be received by either user during the time interval $\Delta_j$ whenever user $u\in[2]$ is scheduled for transmission on layer $q\in[\Qsf]$. The final expression in~\eqref{eq:DTOct04proposed_k_unc} simply says that by time $t^\text{\rm(unc)}_{\pi(j)}$ the fraction of uncoded packets left to be transmitted is proportional to $1-t^\text{\rm(unc)}_{\pi(j)} / t^\text{\rm(unc)}_{q}$ if $\pi(j) <q$ and zero otherwise.
Similarly, we have for $j\in[\Qsf]$:

%%\begin{figure*}[!t]
\begin{align}
&k^\text{(rtx)}_{u}[j]
\notag\\&
   = \Bigg[ k^\text{(rtx)}_{u}[j-1] - \Delta_j \sum_{\ell=1}^{j-1} \Pr[N_u\geq \pi(\ell)] 
\notag\\& 
   +  \sum_{q\in[\Qsf]}\min
%       \hspace*{-1.1cm} 
        \Big(  p, k^\text{\rm(unc)}_{u,q}[j-1] \Big)_{p:=\Delta_j\Pr[A_q=u, \max(N_1,N_2) \geq q, N_u<q]} \Bigg]^+ 
      % \right]^+.
\label{eq:DTOct04proposed_k_rtx}
    \\
  % \right.\notag\\&\left.        
& = \Bigg[  \sum_{\{q:t_q \geq t_{\pi(j)} \}} k_{u,q}^{\text{\rm(rem)}} \frac{t_{\pi(j)}^{\text{\rm(unc)}}}{t_q^{\text{\rm(unc)}}}   
\notag\\& %\hspace*{-1.1cm}
+ \sum_{\{q:t_q < t_{\pi(j)}\}} 
    \left(k_{u,q}^{\text{\rm(rem)}} - (t_{\pi(j)}^{\text{\rm(unc)}} - t_q^{\text{\rm(unc)}})\Pr[N_u \geq q] \right) \Bigg]^+.
\label{eq:DTOct04proposed_k_rtx_final}
\end{align}
%%\hrulefill
%%\end{figure*}

The update equation for $k^\text{(rtx)}_{u}[j]$ in~\eqref{eq:DTOct04proposed_k_rtx} says that the number of coded packets for user $u\in[2]$ can increase or decrease over ``time'' $j\in[\Qsf]$. In particular, at the end of the $j$-th sub-phase, $k^\text{(rtx)}_{u}[j-1]$ is decreased by the number of packets that can be received by user $u\in[2]$ during the time interval $\Delta_j$ on the layers that have already completed their uncoded phase (which is proportional to $\sum_{\ell=1}^{j-1} \Pr[N_u\geq \pi(\ell)]$), or increased by the number of overheard packets during the time interval $\Delta_j$ across any of the layers. The ``min'' in~\eqref{eq:DTOct04proposed_k_rtx} simply says that the number of overheard packets for user $u\in[2]$ on layer $q\in[\Qsf]$ cannot exceed the number of uncoded packets left for transmission at the end of the $(j-1)$-th sub-phase, $k^\text{\rm(unc)}_{u,q}[j-1]$.
The final expression in~\eqref{eq:DTOct04proposed_k_rtx_final} can be derived after some tedious algebra starting form~\eqref{eq:DTOct04proposed_k_rtx}.

At the end of the $\Qsf$-th sub-phase, we have all $k^\text{\rm(unc)}_{u,q}[\Qsf]=0$, but possibly some $k^\text{(rtx)}_{u}[\Qsf]>0$. Therefore, we still have $k^\text{\rm(rem)}_{u}= $ $k^\text{(rtx)}_{u}[\Qsf]$ in~\eqref{eq:schemePrototype} coded packets to deliver to user $u\in[2]$ during Phase2. 
The expression in~\eqref{eq:scheme3DT-Ku} can be obtained after some simple algebra starting from~\eqref{eq:DTOct04proposed_k_rtx_final} with $j=\Qsf$. 
%\nrd{By the time of submission of this paper, we could not find a closed form expression for $k^\text{(rtx)}_{u}[\Qsf]$.} {\red and we are not even sure the analysis is correct; same for the other; question of LLN ...}
\label{eq:DTOct04}
\end{subequations}
\end{IEEEproof}

\section{Numerical Evaluations}
\label{sec:numerical}

\paragraph*{Example~1}
Consider the case of $\Ksf=2$ users and $\Qsf=2$ layers, with $N_1$ independent of $N_2$ and with marginals as in~\cite[eq(29)]{FadingBC}. Without CSIT, the capacity region in Theorem~\ref{thm:Tse-Yates ITA-2011} has three corner points $(R_1,R_2)\in\{ 
(0,1),
(\frac{3}{4},\frac{1}{2}),
(1,0)\},
$
where $1=\mathbb{E}[N_1]=\mathbb{E}[N_2].$
The corner point $(\frac{3}{4},\frac{1}{2})$ is achieved by assigning layer~1 to user~1 and layer~2 to user~2~\cite{FadingBC}.
With COF, it can be shown analytically the outer bound in Theorem~\ref{thm:COFnewOuter} has three corner points $(R_1,R_2)\in\{ 
(0,1),
(\frac{7}{9},\frac{5}{9}),
(1,0)\},
$
and that Theorem~\ref{thm:scheme1} does not achieve the corner point $(\frac{7}{9},\frac{5}{9})$ while Theorem~\ref{thm:scheme2} does (with $R_1 = R_{1,1}$ and $R_2 = R_{2,2}$). This is an example where our bounds are tight. Note that for this channel, one has $t^\text{\rm(unc)}_{1}=t^\text{\rm(unc)}_{2}$, thus there is no issue of ``idle'' slots, which will not be the case for the next example. Notice that COF enlarges the capacity region for this example.

\begin{table}[H]
\caption{\small Joint PMF $\Pr[(N_1,N_2)=(i,j)]$.}
\label{tab:threecorners}
%\vspace*{-.2cm}
\begin{align*}
\begin{array}{| c | c |  c | c | c |}
\hline
&j = 0 &j = 1  &j = 2 & \Pr[N_{1} = i] \\
\hline
i = 0& 0.0497 & 0.2443 & 0.0321 & 0.3261 \\
i = 1& 0.1483 & 0.2251 & 0.1222 & 0.4956 \\
i = 2& 0.0435 & 0.0728 & 0.0620 & 0.1783 \\
\hline
\Pr[N_{2} = j] & 0.2415    &  0.5422   &   0.2163 &  \\
\hline
\end{array} 
\end{align*}
%\vspace*{-.4cm}
\end{table} 

\paragraph*{Example~2}
%%Our achievable regions for the general LPE-BC with COF are relatively easy to describe algorithmically, however deriving a closed form expression is not straightforward. %evaluation of the region is not quite simple, since there are $\Qsf!$ possible permutations of the order of completion time for Phase1. To give a brief view of the performance of our different schemes, let us consider an example  with fading channels correlated and $\Qsf = 2$. Here are the statistics of the channel model
%%Here we numerically evaluate our regions for

\begin{figure}
  \centering
  \includegraphics[width=0.9\columnwidth]{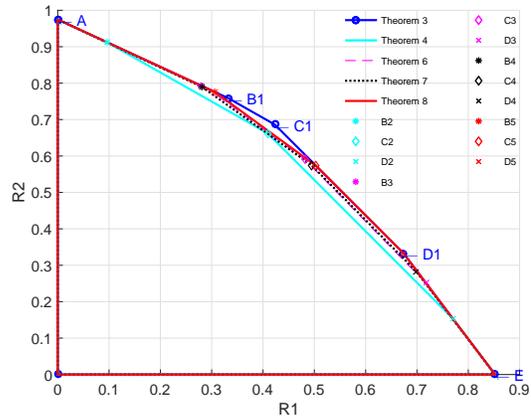}%%{newfig-multicolumnLeg.eps}%{FIG/layer_representation.eps}
  \caption{\small Outer and inner bounds for the channel in Table~\ref{tab:threecorners}.}
  \label{fig:example}
  %\vspace*{-.4cm}
\end{figure}

The inner and outer bound regions for the channel described in Table~\ref{tab:threecorners} are evaluated in Fig~\ref{fig:example}, in which both users have a more reliable look at layer~1 than at layer~2, and the channel states are correlated at each channel use.
%It turns out that this channel is neither degraded, nor less noisy. Plugging all the numbers into \eqref{eq:converse thm:COFnewOuter}, we obtain the outer bound 
%\begin{align}
%& \begin{cases}
%R_{1} + 0.5361R_{2} \leq 0.8522 \\
%1.4100R_{1} + R_{2} \leq 1.2828\\
%R_{1} + 1.2528R_{2} \leq 1.2828 \\
%0.6503R_{1} + R_{2} \leq 0.9748 \label{outer:ieq:R}
%\end{cases}
%\\&= \text{ConvexHull}[A1:(0,0.9748),B1:(0.3326,0.7585), C1: \nonumber \\&(0.4231,0.6862),D1:(0.6739,0.3326), E1:(0.8522,0),(0,0)].
%\end{align}

The outer bound in Theorem~\ref{thm:COFnewOuter} is convex-hull of the following rate points: 
$A=(0,0.9748)$,
$B_1=(0.3326,0.7585)$, 
$C_1=(0.4231,0.6862)$,
$D_1=(0.6739,0.3326)$, 
$E=(0.8522,$ $0)$.
Corner points $A$ and $E$ are always trivially achievable, so we will not list them in the following.
The achievable region in Theorem~\ref{thm:scheme1} has non-trivial corner points:
$B_2=(0.0957,$ $0.9125)$,
$C_2=(0.4091,0.6624)$,
$D_2=(0.7697,0.1540)$.
The achievable region in Theorem~\ref{thm:scheme2} has non-trivial corner points:
$B_3=(0.2779,0.7941)$,
$C_3=(0.4817,0.5903)$,
$D_3= $ $(0.7176, 0.2511)$.
The achievable region in Theorem~\ref{thm:scheme3-SL} has non-trivial corner points:
$B_4=(0.2812,0.7896)$,
$C_4=(0.4943,$ $0.5751)$,
$D_4=(0.6988,0.2827)$.
The achievable region in Theorem~\ref{thm:scheme3-DT} has non-trivial corner points:
$B_5=(0.3069,0.7752)$,
$C_5=(0.5035,0.5729)$,
$D_5=(0.6739,$ $0.3326)$.
It is not easy to tell the difference among the various achievable regions with the naked eye from Fig~\ref{fig:example}, but the order of inclusion, from the smallest to the largest region is,
 Theorem~\ref{thm:scheme1},
 Theorem~\ref{thm:scheme2},
 Theorem~\ref{thm:scheme3-SL}, 
 Theorem~\ref{thm:scheme3-DT}, and finally the outer bound in
Theorem~\ref{thm:COFnewOuter}.
It is noticed that Theorem~\ref{thm:scheme3-DT} achieves one of the corner points ($D_1$) of the outer bound in Theorem~\ref{thm:COFnewOuter}.
An interesting observation from the numerical optimization for this example is that at the corner points either $k_{1,q}=0$ or $k_{2,q}=0$ in the various achievable regions across layers (i.e., a layer is assigned to one user only -- as it was the case in Example~1), with the only exception of C-points; for the C-points, the `more reliable'  layer~1 is shared by both users. We also remark from Fig~\ref{fig:example} that the inner and outer bounds are the furthest apart around C-points. Why this is the case is subject of current investigation.

%{\blue
%SL: The parameters of each achievable corner point are listed as below:(some of the achievable rates have multiple schedule methods.)
%\begin{align*}\tiny
%\begin{array}{lllll}
%B2:
%&R_{11} = 0,      &R_{12} =0.2812, &R_{21} = 0.7896, &R_{22} = 0;\\
%C2:
%&R_{11} = 0.4943, &R_{12} =0,      &R_{21} = 0.2827, &R_{22} = 0.2924;\\ %INTERESTING
%D2:
%&R_{11} = 0.6988, &R_{12} =0,      &R_{21} = 0,      &R_{22} = 0.2827;\\
%B3: 
%&R_{11} = 0,      &R_{12} =0.2991, &R_{21} = 0.7798, &R_{22} = 0;\\
%C3:
%&R_{11} = 0.4435, &R_{12} =0.0428,      &R_{21} = 0.3022, &R_{22} = 0.2827;\\%INTERESTING
%D3:
%&R_{11} = 0.6992, &R_{12} =0,      &R_{21} = 0.0010,      &R_{22} = 0.2841;\\
%\end{array}
%\end{align*}
%}
%{\red
%DT: All cases except C2 and C3 may be easy to verify analytically, these are the cases where a layer is given to one user only.
%More interesting are C2 and C3 where layer1 is shared among the two users; in these case it would be useful to know the valued of all the parameters in the ach.regions and also run the algorithm to see if the formulas we write do make sense at least for this example. 
%}
%{\blue SL: All the parameters of B2,C2,D2, B3, C3, D3 have been checked. They can be achieved and also make sense for this example}

\section{Conclusions}
\label{sec:concl}
This paper derived inner and outer bounds for the LPE-BC with COF. The studied LPE-BC extends the classical (single-layer) binary erasure BC and has can be connected to the Gaussian fading BC. Our inner bounds make use of network coded retransmissions when the sender, through COF, realizes that a packet has been received only by unintended users.  What this work shows is the necessity of network coding across users (a key element also for the single-layer binary erasure BC with COF) and across layers. Analytical and numerical examples confirm that our bounds can be tight for some channel parameters. Future work includes determining for which channel parameters the presented schemes are optimal, deriving new strategies for the remaining cases, extending the analysis to more than two users, and ultimately derive schemes for the Gaussian noise case.

%\section*{Acknowledgment}

\bibliographystyle{IEEEtran}
\bibliography{refs-ND}%refs}%,

\end{document}